\documentclass[twocolumn,prl,nopacs,floatfix]{revtex4}
\usepackage{graphicx}
\usepackage{epsfig}
\usepackage{rotating}
\usepackage{amsmath}
\usepackage{amsfonts}
\usepackage{amssymb}
\usepackage{enumerate}
\usepackage{longtable}
\usepackage{dcolumn}
\newcolumntype{.}{D{.}{.}{-1}}
\usepackage{tabularx}
\usepackage{dcolumn}
\usepackage{bm}
\newcommand{\chiSG}{\chi_{_{SG}}}
\newcommand{\av}{_\mathrm{av}}
\begin{document}

\title{Critical and Griffiths-McCoy singularities in quantum Ising spin-glasses on d-dimensional
hypercubic lattices: A series expansion study}

\author{R. R. P. Singh}
\affiliation{University of California Davis, CA 95616, USA}

\author{A. P. Young}
\affiliation{University of California Santa Cruz, CA 95064, USA}

\date{\rm\today}

\begin{abstract}
We study the $\pm J$ transverse-field Ising spin glass model at zero
temperature on d-dimensional hypercubic lattices and in the
Sherrington-Kirkpatrick (SK) model, by
series expansions around the strong field limit. In the SK model and in
high-dimensions our calculated critical properties are in excellent agreement
with the exact mean-field results, surprisingly
even down to dimension $d = 6$ which is below the
upper critical dimension of $d=8$.  In contrast, in lower dimensions we
find a rich singular behavior consisting of critical and Griffiths-McCoy
singularities. The divergence of the equal-time structure factor allows us to
locate the critical coupling where the correlation length diverges, implying
the onset of a thermodynamic phase transition. We find that the spin-glass
susceptibility as well as various power-moments of the local susceptibility
become singular in the paramagnetic phase \textit{before} the critical point. 
Griffiths-McCoy singularities are very strong in two-dimensions but decrease
rapidly as the dimension increases. We present
evidence that high enough powers of the local susceptibility may become singular
at the pure-system critical point.

\end{abstract}


\maketitle

The combination of quantum mechanics and disorder leads to rich behavior at
and near zero temperature quantum critical points (QCPs). For example, the one
dimensional random transverse field Ising model has a QCP in which average and
typical correlation functions have different critical
exponents~\cite{fisher:92,fisher:95}, and the time-dependence is described by
activated dynamical scaling, in which the log of the relaxation time is
proportional to a power of the correlation length $\xi$, rather than
conventional dynamical scaling in which the relaxation time itself is
proportional to $\xi^z$, where $z$ is dynamical exponent. Distributions of
several quantities are very broad at the quantum critical point (QCP) so QCP's
with these features are said to be of the ``infinite-randomness'' type.
It has been proposed~\cite{fisher:99} that infinite-randomness QCP's can occur in
dimension higher than $1$. It is also proposed~\cite{fisher:99} that the 
infinite-randomness QCP can occur in spin glasses,
on the grounds that frustration is irrelevant since the distribution of
renormalized interactions (as one perform renormalization group
transformations) is so broad that only the largest one matters. 

In addition,
singularities can occur in the paramagnet phase in the region where the
corresponding \textit{non-random} system would be ordered. This was first
pointed out for classical systems by Griffiths~\cite{griffiths:69}, though the
singularities turn out to be unobservably weak in that case~\cite{harris:75}.
However, these singularities are much stronger in the quantum case, as first
shown by McCoy~\cite{mccoy:69,mccoy:69b}, and can lead to power-law
singularities in \textit{local} quantities in part of the paramagnetic phase. For
quantum problems we will refer to these effects as Griffiths-McCoy (GM)
singularities. For a review see Ref.~\cite{vojta:10}, and for recent
experimental observations of GM singularities see Ref.~\cite{wang:16}. In the
quantum paramagnetic phase in the limit as $T \to 0$ GM singularities are
characterized by a dynamical exponent $z'$ which varies as the QCP is
approached. For infinite-randomness QCP's, $z' \to \infty$ as the QCP is
approached~\cite{fisher:99,vojta:10}.

In this paper we study the QCP and GM singularities in quantum spin glasses.
The infinite-range version, the Sherrington-Kirkpatrick
(SK)~\cite{sherrington:75} in a transverse field, has been studied in
detail~\cite{ye:93,huse:93,read:95}, and the mean field behavior determined.
There have also been quantum Monte Carlo (QMC) studies in dimension $d$ equal
to two~\cite{rieger:94} and three~\cite{guo:94}. In this paper we study
quantum spin glasses using series expansions at $T=0$, in which we expand away
from the high transverse-field limit.  We feel that the series expansion method is
complementary to QMC simulations and has certain advantages including: (i) we
study the \textit{whole range} of dimensions from $d=2$ to the SK model
(which is effectively
infinite-$d$), (ii) we work
at strictly zero temperature whereas in QMC one has to extrapolate to $T=0$
using a rather complicated anisotropic finite-size scaling
procedure~\cite{rieger:94,guo:94}, and (iii) averaging over bond disorder is
done exactly. Like QMC we can see GM singularities (we think our work
is the first time these singularities have been seen using series methods),
and also go beyond simple averages of local quantities, in our case by
computing moments up to high order.

Our main conclusions are as follows.
We show systematically how the strength of GM singularities diminish rapidly as the
dimension increases above two, vanishing, as expected, for the SK model. In
two dimensions, where GM singularities are strongest, our results make
plausible the expectation that GM singularities persist in the paramagnetic
phase all the way to the critical point of the pure system. We find that
critical behavior close to that of mean field theory persists \textit{below} the upper
critical dimension, $d_u=8$~\cite{read:95}, down to $d = 6$, which is
surprising since renormalization group finds no perturbative fixed point below
$d=8$~\cite{read:95}. Our results in two and three dimensions agree very well
with earlier work~\cite{rieger:94,guo:94}.

We consider the Hamiltonian
\begin{equation}
\mathcal{H} = - h^T \sum_{i=1}^N \sigma_i^x
-\sum_{\langle i, j\rangle} J_{ij} \sigma_i^z \sigma_j^z
\, ,
\end{equation}
where the $\sigma_i^\alpha$ are Pauli spin operators and $h^T$ is the
transverse field. The
interactions $J_{ij}$ are \textit{quenched} random variables with a bimodal
distribution.
The $N$ spins either lie on a hypercubic lattice, in which
case the
interactions are between nearest-neighbors and take values $\pm J$ with equal
probability,
or correspond to the
Sherrington-Kirkpatrick (SK)~\cite{sherrington:75}
model in which case there is no lattice
structure, every spin interacts with every other spin,
and
$J_{ij} = \pm J / \sqrt{N}$. We choose a bimodal distribution 
because the series can be worked out much more
efficiently for this case than for a general distribution~\cite{singh:17c}. 
We will also add longitudinal fields $h_i$, coupling to $\sigma_i^z$ to define
the spin glass susceptibility, and set them to zero afterwards, see
Eqs.~\eqref{chiSG} and \eqref{Eh} below.

The zero temperature quantities we calculate are:
\begin{itemize}
\item
The zero-wavevector, equal-time structure factor defined as
\begin{equation}
S(0) = {1\over N} \sum_{i,j} [\langle 0|\sigma_i^z \sigma_j^z|0\rangle^2]\av,
\end{equation}
where the state $|0\rangle$ is the ground state of the system, and the average
$[\cdots]\av$ refers to
disorder average over the quenched random bonds.
\item
The spin-glass susceptibility defined as
\begin{equation}
\chiSG={1\over N}\sum_{i,j}\ [\chi_{ij}^2]\av
\label{chiSG}
\end{equation}
where the ground state energy $E(\{h_i\})$ in the presence of infinitesimal local
longitudinal fields $h_i$ defines the
local susceptibilities $\chi_{ij}$ by the relation
\begin{equation}
E(\{h_i\}) = E_0 - {1\over 2}\sum_{i,j}\chi_{ij}\ h_i\ h_j \, .
\label{Eh}
\end{equation}
\item
The moments of the local susceptibility defined as
\begin{equation}
\chi_m={1\over N}\sum_i\ [\chi_{ii}^m]\av \, .
\end{equation}
\end{itemize}

\begin{table}[ht]
\caption{\label{table}Estimates of points of singularity and exponents in
various dimensions and the SK model. Note that $1/x_c\equiv
(h^T_c/J)^2$. 
We anticipate that the singularity found for the equal time
structure factor $S(0)$ is the critical singularity and so has exponent
$\gamma - 2 z \nu$.  If the QCP is of the infinite-randomness type, then $z$
and $\gamma$ are infinite but the combination $\gamma - 2 z \nu$ is presumably
finite. For $\chiSG$ the critical exponent is
$\gamma$, but for low dimensions the susceptibility singularity is clearly at
a larger value of $1/x_c$ than the critical singularity determined from
$S(0)$, i.e.~it is in the paramagnetic phase. Consequently, the series is finding a GM
singularity for $\chiSG$ rather than the critical singularity, so we denote the exponent
by $\lambda$ rather than $\gamma$. For the SK model GM singularities do not
occur, so $\lambda = \gamma$ and the exact values are
$\gamma = 1/2, \gamma - 2 z \nu = -1/2$~\cite{ye:93,huse:93,read:95}, with
logarithmic corrections for $\chiSG$ as discussed in the text.
The value of $(h^T_c/J)^2$ is estimated to be
$2.268$ in Ref.~\cite{yamamoto:87} and $2.28\pm0.03$ in Ref.~\cite{young:17b}.
In $d= 6$ and $8$, GM singularities are very
weak, since  the two values of $x_c$ are almost the same, we expect
that the series for $\chiSG$ gives the critical singularity in those cases too.
For the SK model, and for $d = 6$ and
$8$, we show results for $\chiSG$ both with and without the mean-field
log correction. No log correction is applied to $S(0)$ so the results
for this quantity are the same in both rows.
It is
curious that the difference between the exponents for $\chiSG$ and $S(0)$ is
close to $1$ for all the  models studied.
}
\begin{tabular}{|c|.|c|.|c|}
\hline
Model & \multicolumn{2}{c|}{$\chiSG$} & \multicolumn{2}{c|}{$S(0)$} \\
\hline
& \multicolumn{1}{|c|}{$ 1/x_c$}& $ \lambda $ & \multicolumn{1}{|c|}{$1/x_c$}& $\ \gamma - 2 z \nu$\  \\
\hline \hline
SK           &   2.267  &  0.578    &  2.265 &   -0.509           \\
SK (log)     & 2.268    & 0.486     &  2.265 &  -0.509            \\
SK \cite{ye:93,huse:93,read:95,yamamoto:87}   &2.268   & $1/2$
&2.268       & $-1/2    $ \\
SK \cite{young:17b} &2.28(3)&  &2.28(3)&  \\
\hline
8d (log) & 32.92  & 0.475     & 32.77 &  -0.515  \\
8d       & 32.83  & 0.606     & 32.77 &  -0.515  \\
\hline
6d (log) & 23.75  & 0.489     & 23.60 &  -0.513  \\
6d       & 23.68  & 0.628     & 23.60 &  -0.513  \\
\hline
4d & 14.46  & 0.697     & 14.23 &  -0.441  \\
3d & 9.791 & 0.796     & 9.411 &  -0.300\\
2d & 5.045 & 1.281     & 4.521 &   0.284 \\
\hline
\end{tabular}
\end{table}

\begin{figure*}
\includegraphics[scale=0.33]{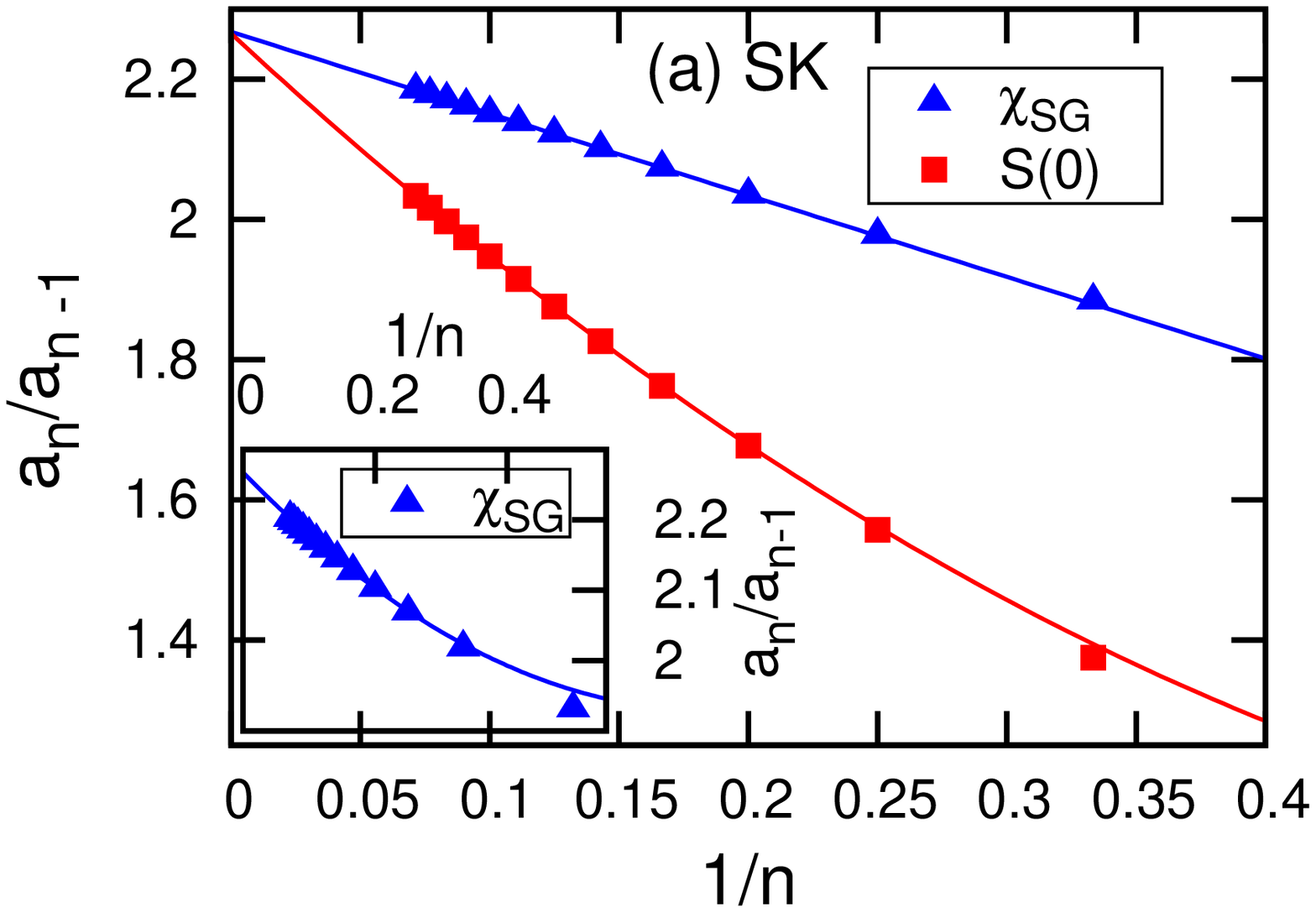}
\includegraphics[scale=0.33]{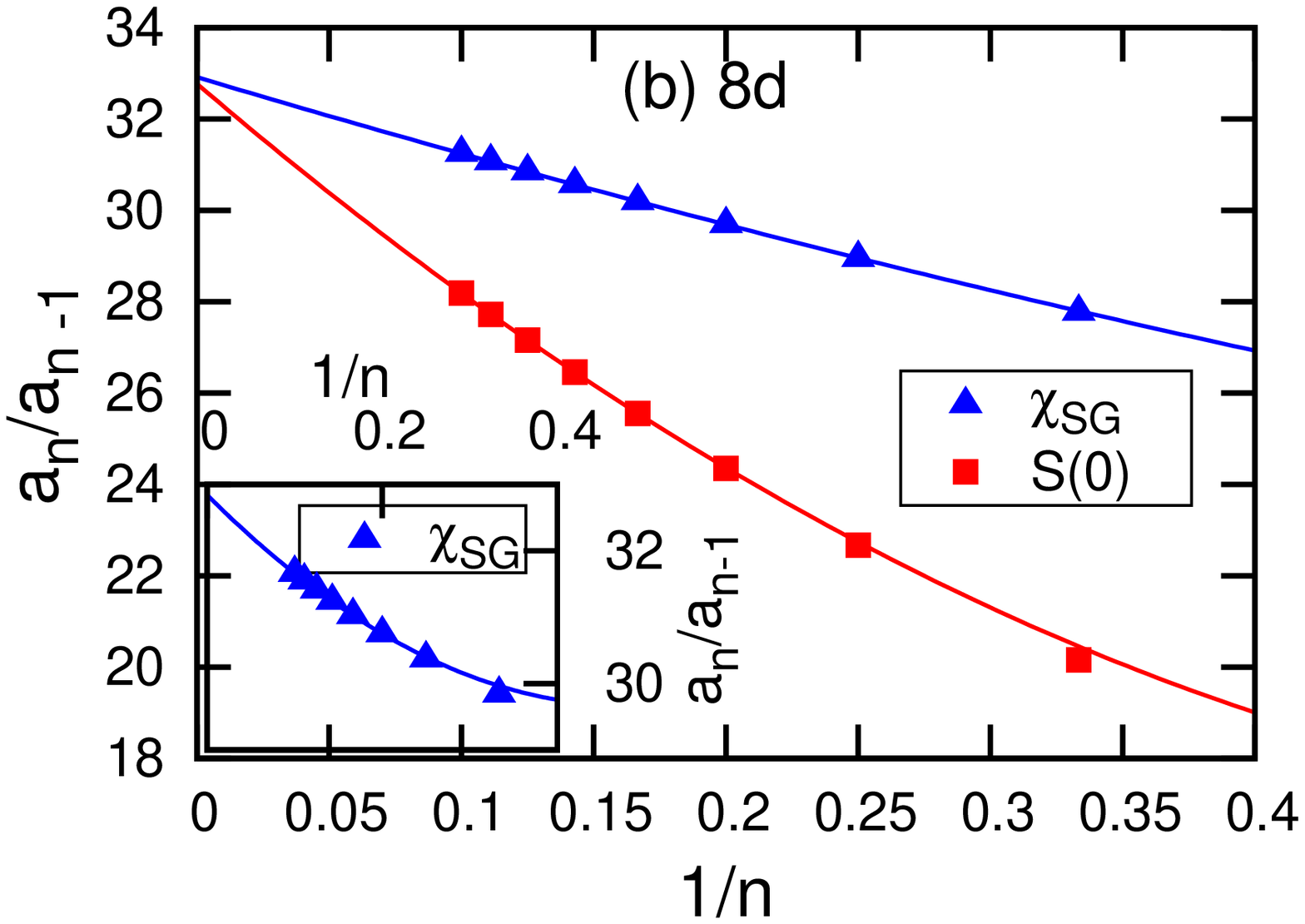}
\includegraphics[scale=0.33]{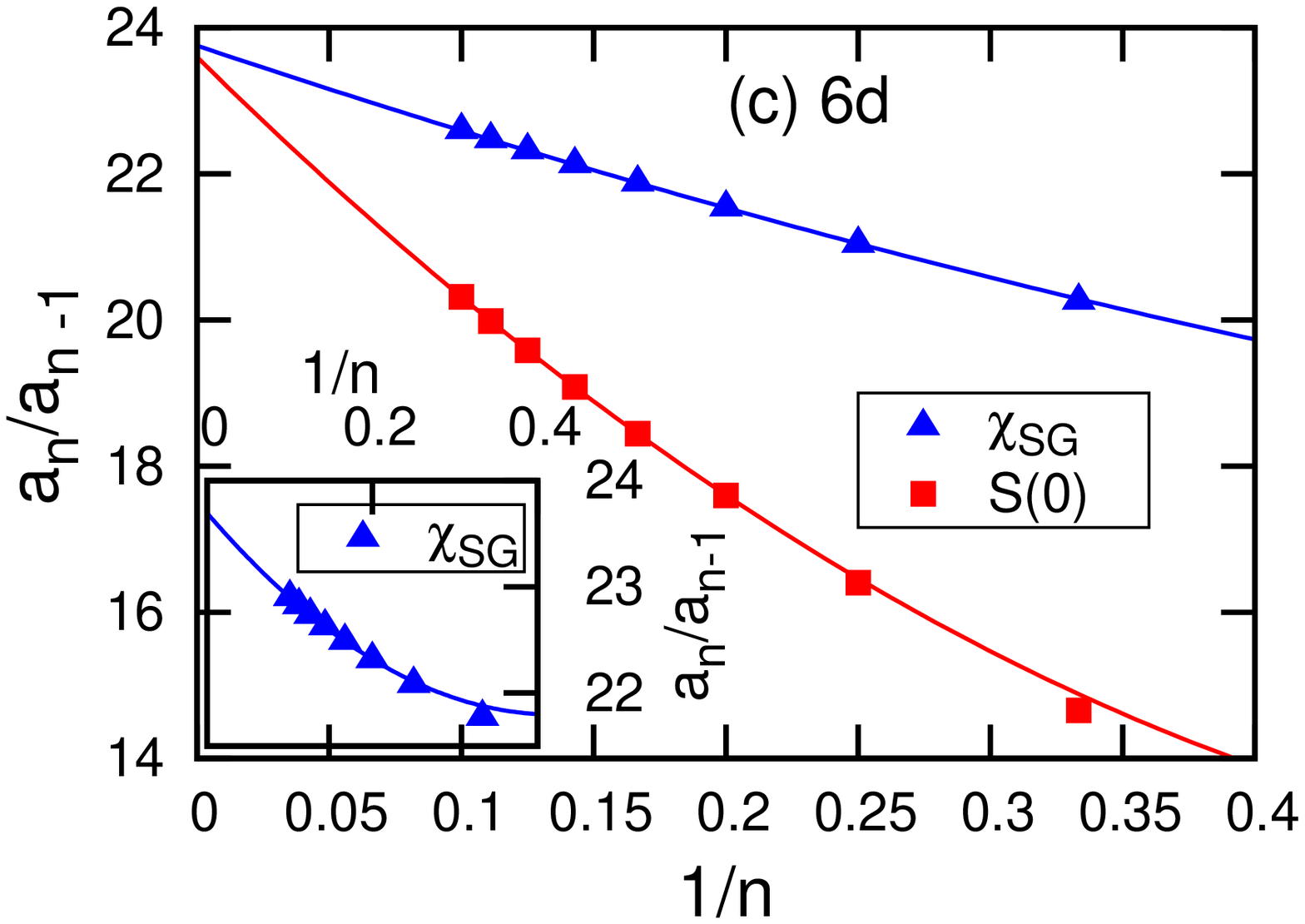}
\caption{Ratio plots for (a) the SK model, (b) $d=8$, and (c) $d=6$. In all cases
the differences in the intercept, i.e.~the values of $1/x_c$, between
results for $S(0)$ and $\chiSG$, is very small or zero indicating
that GM singularities are very small or non-existent. Thus it is plausible
that the singularity exponent for $\chiSG$ is the \textit{critical}
exponent $\gamma$ for these figures. The fits are quadratic except for $\chiSG$ for the SK
model where we used a linear fit.
The parameters of the fits are
given in Table~\ref{table}. In all three figures, the series for $\chiSG$ had the mean-field
log factor~\cite{ye:93,huse:93,read:95} incorporated.
The insets show the ratios for $\chiSG$ series \textit{without} the
log factor. The greater curvature is apparent.
\label{figs_highd}
}
\end{figure*}

We use the linked cluster method to generate the series~\cite{gelfand:90} and
discuss
the details of the computational method
elsewhere~\cite{singh:17c}. We expand away from the trivial paramagnetic state
with $J=0$, so the expansion parameter is 
\begin{equation}
x = (J / h^T)^2 .
\end{equation}
The series are obtained to order $14$ for the SK
model and for $d=2$ and $3$, and to order $10$ in higher dimensions.  The
series coefficients can be found in the source material on the arXiv.  Throughout this
paper we generically refer to the coefficients of the series expansions as
$a_n$, meaning that the series is of the form
\begin{equation}
Q = \sum_n a_n x^n \, .
\end{equation}
We find that, in contrast to
classical spin-glasses~\cite{singh:86,daboul:04},
the series for the quantum systems are surprisingly
well behaved.
Most of our analysis is based on the
simple ratio method, although, we have checked that d-log Pade analysis gives
answers consistent with them.
If the series has a simple power-law
variation, $Q \propto (x - x_c)^{-\lambda}$, then the ratios satisfy
\begin{equation}
r_n \equiv {a_n \over a_{n-1}} = {1 \over x_c}\left(1 + {\lambda -1 \over
n}\right).
\end{equation}
Hence, in a plot of the ratio $r_n$ against $1/n$, the intercept gives $1/x_c$ and the
slope for $1/n \to 0$ gives $(\lambda -1)/x_c$. We will do linear and
quadratic fits to our data to extract $x_c$ and the exponent.

Griffiths-McCoy (GM) singularities occur in a quantum disordered system at
low-$T$ in the region between the critical point of the system and the
critical point of the corresponding pure system. In this range, there are
regions of the sample which are non-disordered and so are ``locally in the
ordered, symmetry-broken state''. The very slow tunneling between
between the symmetry-broken states leads to power-law
singularities~\cite{thill:95,guo:96,rieger:96b,vojta:10} in the paramagnetic phase,
coming from purely local physics, namely
a distribution of local relaxation times which extends up to very high values.

Since $\chiSG$ is the divergent response function for this problem, its
\textit{critical} exponent is defined to be~\cite{read:95} $\gamma$, i.e.
\begin{equation}
\chiSG \propto (x_c - x)^{-\gamma}.
\end{equation}
However, it is important to stress that, because of GM singularities,
the exponent determined
in the series is usually different from $\gamma$, and shall generally call it
$\lambda$, see Table~\ref{table}.
The equal time structure factor $S(0)$ does not have the two
time-integrals present in $\chiSG$. Since the dynamic exponent is $z$ and the
correlation exponent is $\nu$, the critical behavior of $S(0)$ is 
\begin{equation}
S(0) \propto (x_c - x)^{-(\gamma - 2 z \nu)}.
\end{equation}
\label{S0}
Because there are no time integrals in $S(0)$ we expect that GM singularities
will not occur for this quantity, and any classical-like
Griffiths~\cite{griffiths:69} singularities will be unobservably
weak~\cite{harris:75}.
If the QCP is of the infinite-randomness type, then $\gamma$ and $z$ will be
infinite, though presumably the combination $\gamma - 2 z \nu$ will be finite
since it describes the critical behavior of an \textit{equal time} quantity
$S(0)$.

\begin{figure*}
\includegraphics[scale=0.33]{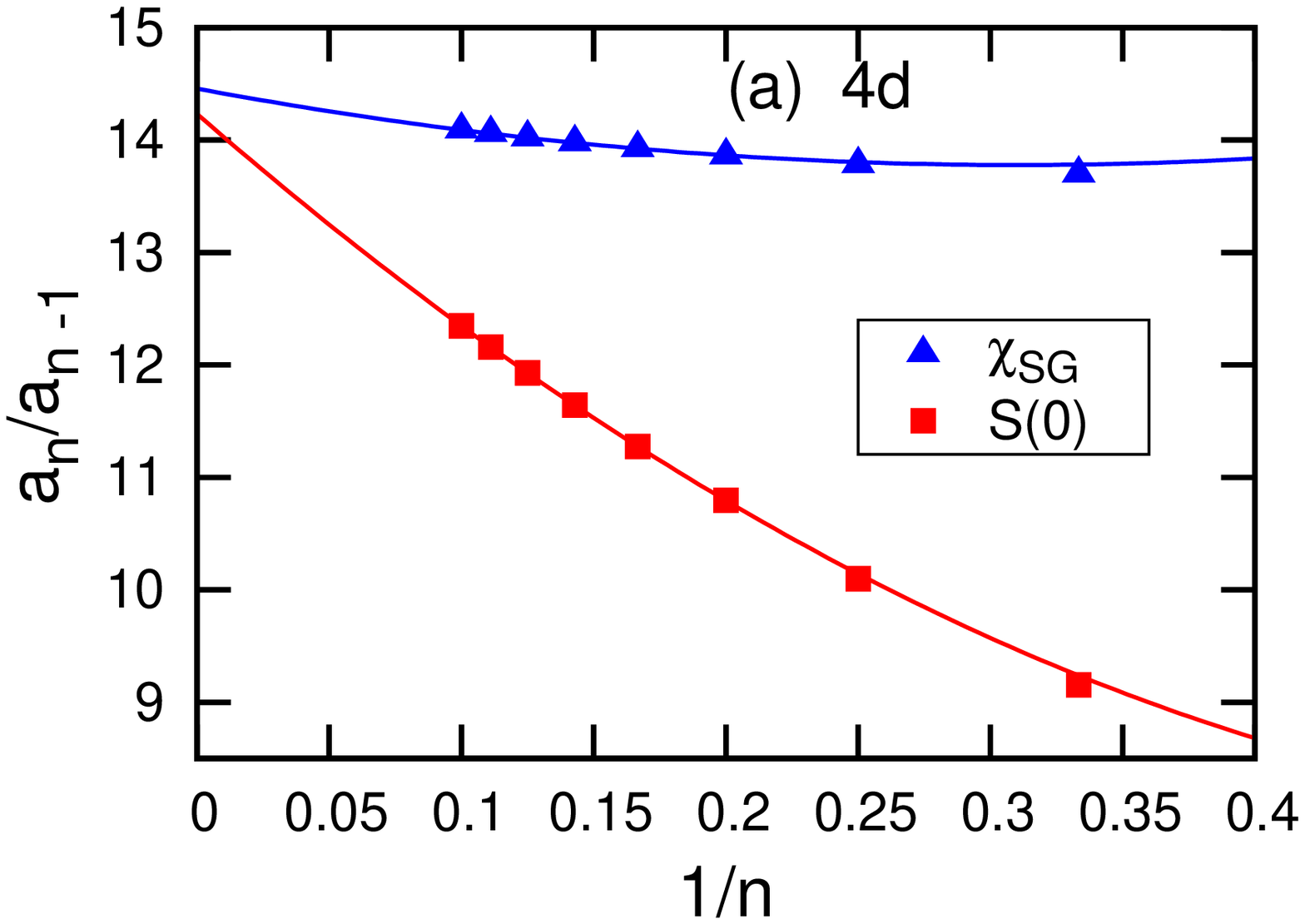}
\includegraphics[scale=0.33]{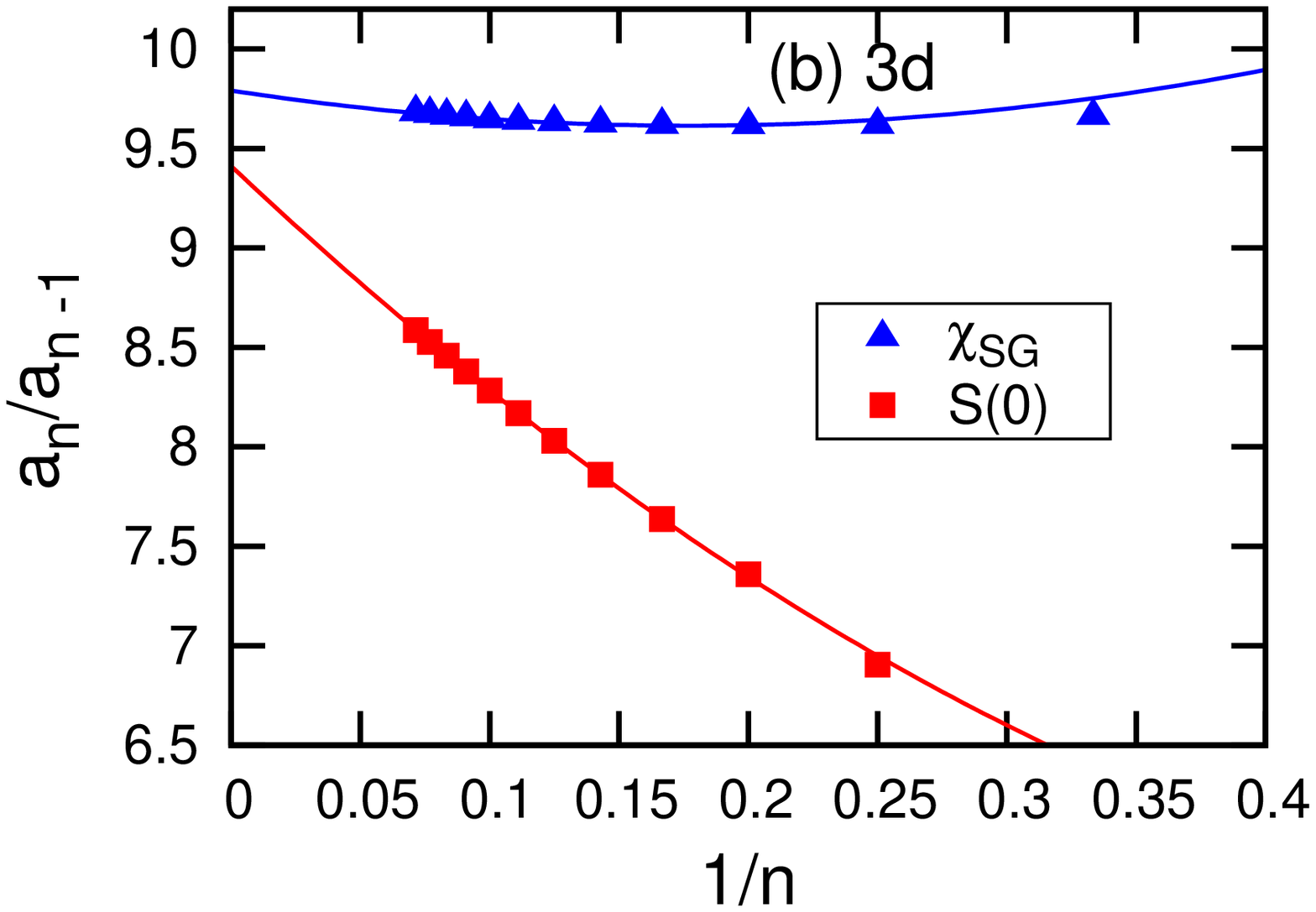}
\includegraphics[scale=0.33]{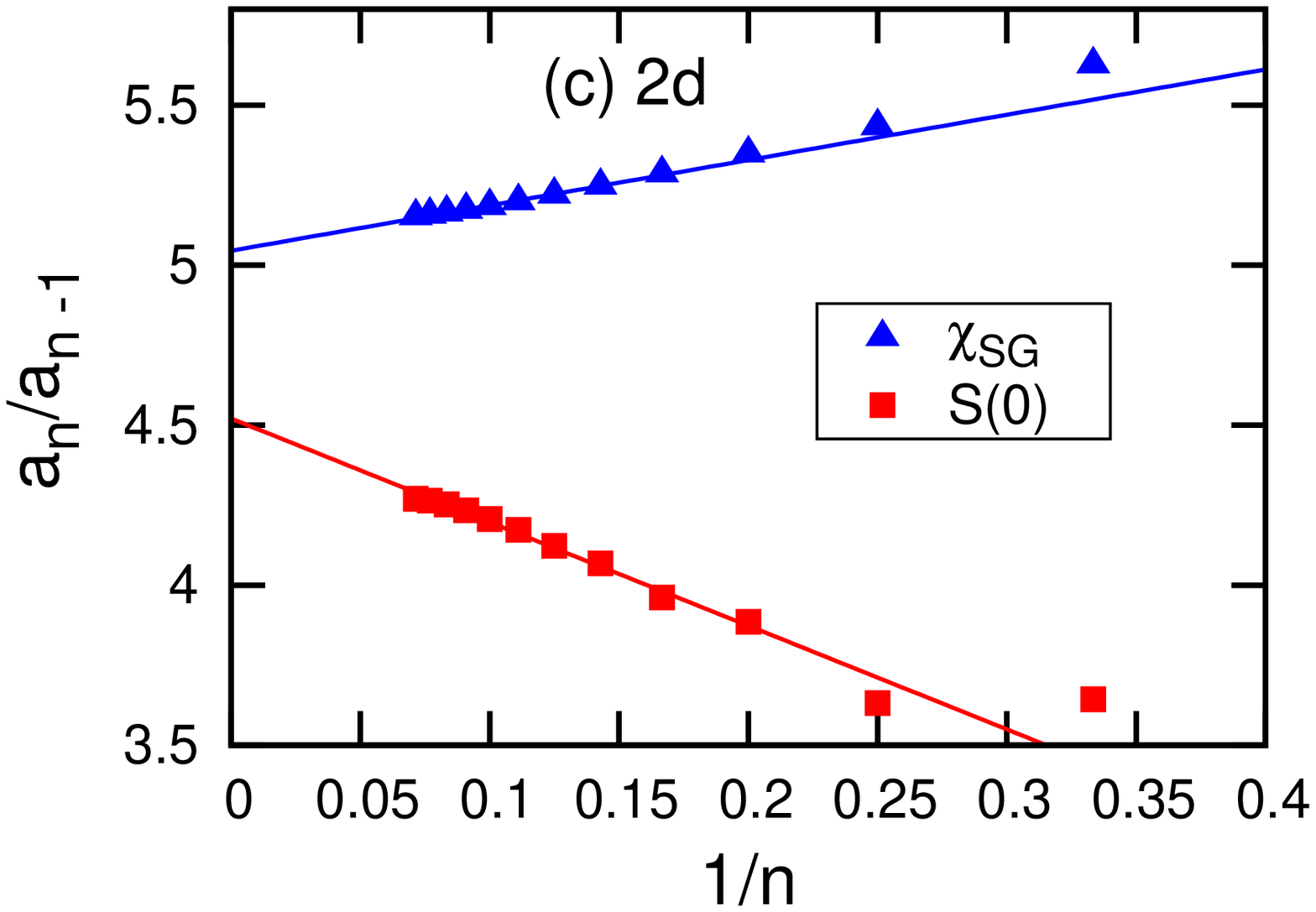}
\caption{Ratio plots for (a) $d = 4$, (b) $d=3$, and (c) $d=2$. In all cases
there is a difference in the intercept, i.e.~the value of $1/x_c$, between
results for $S(0)$ and $\chiSG$, indicating that $\chiSG$
diverges in the paramagnetic phase \textit{before} the QCP is reached, i.e.~the
singularity in $\chiSG$
corresponds to GM singularities, not critical singularities. Note that the
difference in $x_c$ values is very large in 2d but rapidly decreases with
increasing dimension. The fits are linear for $d=2$ and quadratic for $d =
3$ and $4$. The parameters of the fits are given in
Table~\ref{table}.
\label{figs_lowd}
}
\end{figure*}

We now discuss our results. Figures \ref{figs_highd} and \ref{figs_lowd}
show the ratios $r_n$ for
each model. The reader should also refer to Table~\ref{table} for
values of the points of singularity and exponents.

We begin with the results for the SK model shown in Fig.~\ref{figs_highd}(a).
The critical point obtained,
$1/x_c=2.268$, is in excellent agreement with other
studies~\cite{yamamoto:87,young:17b}. The analytic
prediction~\cite{ye:93,huse:93,read:95} is $\gamma= 1/2$ with a log-correction.
To account for the log,
we divide the series by $[ -(1/t) \log(1-t) ]^{1/2}$, where
$t=x/x_c$. Ratios of the resulting series give $\gamma=0.49$, in excellent
agreement with the exact result.
Without taking logarithms into account, the exponent $\gamma$ is
estimated too high as $0.58$. Other predictions are~\cite{read:95} $z = 2, \nu
= 1/4$, so $\gamma - 2 z \nu = -1/2$ (the exponent for $S(0)$).
Our result for this is $-0.51$ again in good agreement. The critical points
for $\chiSG$ and $S(0)$ agree with each other to high precision indicating
that there are no GM singularities in the SK mode, as expected.

Results of the ratio analysis of the series in $d=8$ and $d=6$ are shown in
Fig.~\ref{figs_highd}(b)-(c). Curiously the results for $\chiSG$ work better,
in the sense that the ratio plot is closer to a straight line, if one
includes the \textit{same} mean-field log-correction as for the SK model. Of
course, even if there are log corrections there is no \textit{a priori} reason
to assume that they have the same form as in mean field theory. Including
this correction the exponent for $\chiSG$ is very close to the mean field
value of $1/2$. The exponent for $S(0)$ (for which no log-correction is
performed) is  also close to the mean field prediction of $-1/2$. There is very
little difference in the critical points for $\chiSG$ and $S(0)$ indicating
that GM singularities, if present, are very weak. It is surprising that the
same near-mean-field-like behavior is found in $d=6$ as well as $d=8$ since $d
= 8$ is the upper critical dimension for this problem and no
perturbative fixed point is found~\cite{read:95} below $d=8$. One might
therefore
expect a dramatic change in critical behavior in going below $d =8$, but this is not
what we find.

In $d=4$, see Fig.~\ref{figs_lowd}(a),
there are clear deviations from mean field exponents, and a clear, though small,
difference
between the critical points for $\chiSG$ and $S(0)$ indicating the presence of
rather weak GM singularities.

\begin{figure}
\begin{center}
 \includegraphics[angle=0,width=0.9\columnwidth]{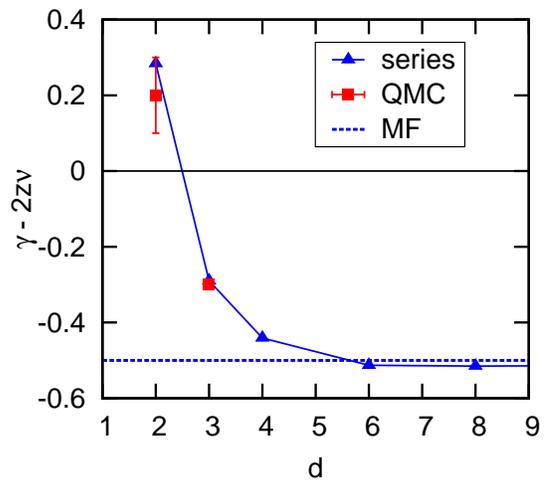}
\caption{
The critical exponent $\gamma - 2 z \nu$ for $S(0)$ as a function of
dimension. It is seen that no apparent change occurs at the upper
critical dimension
of $d=8$~\cite{read:95}, and that our results in $d=2$ and $3$ agree well with
QMC simulations of Ref.~\cite{rieger:94} and \cite{guo:94} respectively.
\textit{Note:} this figure is not included in the published version.
\label{fig_exponents}}
\end{center}
\end{figure}

Comparing the results for $d=4$ with those for $d=3$ and $2$ in
Fig.~\ref{figs_lowd}(b) and (c), we see that the strength of GM singularites
increases considerably with decreasing dimension. The same conclusion follows
from comparing
QMC results in $d=3$~\cite{guo:96} with those in $d=2$~\cite{rieger:96b}. For
the critical singularity of $S(0)$ in $d=3$
we find an exponent $\gamma - 2 z \nu =
-0.30$ which is in excellent agreement with the QMC calculations of
Ref.~\cite{guo:94} who obtain $-0.3$, which we deduce from their results $z
\simeq 1.3, 1/\nu \simeq 1.3, \eta + z \simeq 1.1$~\cite{eta_problem} and the
scaling relation $\gamma / \nu = 2 -\eta$. There is also good agreement
in $d=2$ between our value of $0.28$ for the exponenent for $S(0)$ and the QMC
value~\cite{rieger:94} of $0.2 \pm 0.1$ which we deduce from their value of
$(\gamma - 2 z \nu)/\nu = 0.2 \pm 0.1$ in their Fig.~4, and $\nu = 1.0 \pm
0.1$. Our results for the exponent for $S(0)$ are summarized in
Fig.~\ref{fig_exponents}.

\begin{figure}
\begin{center}
 \includegraphics[angle=0,width=0.9\columnwidth]{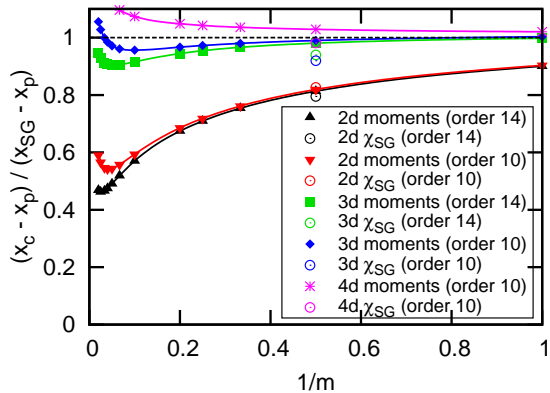}
\caption{
Location of the singularity for the $m$-th moment of the local susceptibility
in $d=2,3$ and $4$, plotted as a function of $1/m$. Here $x_p$ is the critical
point of the pure system, and $x_{SG}$ is the spin glass
critical point as determined from equal-time structure factor,
so the $y$-axis is scaled such that the pure system critical point
corresponds to $y=0$ and the spin glass critical point to
$y=1$. The spin glass phase corresponds to $y > 1$ and the region of GM
singularities to $0 < y < 1$.
\label{fig_moments}}
\end{center}
\end{figure}

We now study the GM singularities in more detail, focussing on the
\textit{local} susceptibility $\chi_{ii}$. According to the standard
picture~\cite{vojta:10,thill:95,guo:96,rieger:96b}, GM singularities occur
because $\chi_{ii}$ is a random quantity, with a broad distribution extending
out to very large values. Although we can't compute the distribution of
$\chi_{ii}$ directly we can get information on it \textit{indirectly} by
computing the series for moments of it, up to high order. 
The results are summarized in Fig.~\ref{fig_moments}. The $y$-axis is defined
such that the critical value of $x$ for the spin glass problem is at $y=1$ and
the critical value for the pure ferromagnet (i.e.~all interactions equal to
$J$) is at $y=0$.

Consider first the results for $d=2$ shown in Fig.~\ref{fig_moments}.
We see that the higher moments have a
singularity further and further away from the spin glass critical point (which
corresponds to $y=1$).
For large values of the order of the moment $m$, the 14-th order series lies
below the 10-th order series. It is therefore plausible that, for an
infinitely long series, the singularity approaches the pure system critical
point (i.e.~$y=0$) for $m \to \infty$.

Also shown in Fig.~\ref{fig_moments}, by circles,
are the locations of the divergence of the
spin-glass susceptibility. This quantity has two time integrals and so we put
these points at $m=2$. In $d=2$ the $\chiSG$ singularities agree very well
with the singularities in the local-$\chi$, confirming that the singularity
found in $\chiSG$ is a (local) GM singularity, not the critical singularity.

In $d = 3$, there seems to be a difference in Fig.~\ref{fig_moments}
between the local-$\chi$ and
$\chiSG$ results, but notice the opposite trends in the data between the
14-term and 10-term series, so it is plausible that the two quantities would
be singular at the same point in an infinitely-long series.

In $d=4$, the GM
singularities are sufficiently weak that the local-$\chi$ does not show a
singularity at the QCP or in the paramagnetic phase, at least with a 10-term
series. We expect that the singularity would be at the same location as that
of $\chiSG$ (which is just in the paramagnetic phase) for an infinitely long series.

To conclude, we
have shown systematically how the strength of GM singularities diminish rapidly
as the
dimension increases above two, vanishing, as expected, for the SK model. In
two dimensions, where GM singularities are strongest, our results make
plausible the expectation that GM singularities persist in the paramagnetic
phase all the way to the critical point of the pure system. We find that
critical behavior close to that of mean field theory persists \textit{below}
the upper
critical dimension, $d_u=8$~\cite{read:95}, down to $d = 6$, which is
surprising since renormalization group finds no perturbative fixed point below
$d=8$~\cite{read:95}. Our results in two and three dimensions agree very well
with earlier work~\cite{rieger:94,guo:94}.

Since the series for $\chiSG$ sees GM singularities rather than 
critical singularities we can not determine whether or not the QCP is of the
infinite-randomness type (for which $\gamma$ and $z$ are infinite, though
$\gamma - 2 z \nu$ is finite). Recent numerical
simulations in two dimensions~\cite{matoz-fernandez:16} are argued to support the
infinite-randomness scenario, though it seems to us that conventional critical
behavior fits the data about as well. As the dimension increases, the effects
of GM singularities become much weaker than in $d=2$, so we conclude that
if the infinite-randomness scenario occurs at all for $d > 2$, it must manifest itself
only over a very small region around the quantum critical point.

{\it Acknowledgments:}
One of us (APY) would like to thank the hospitality of the Indian Institute of
Science, Bangalore and the support of a DST Centenary Chair Professorship.
He is particularly grateful
for stimulating discussions with H.~Krishnamurthy which initiated this
project. We would like to thank Thomas Vojta for informing us of recent
experimental work on GM singularities and for an informative correspondence on
infinite-randomness quantum critical points. We would like to thank
D.~A.~Matoz-Fernandez for bringing Ref.~\cite{matoz-fernandez:16} to our
attention.
The work of RRPS is supported in part by US NSF grant
number DMR-1306048.

\bibliography{refs,comments}

\begin{thebibliography}{26}
\bibitem[{eta()}]{eta_problem}
\bibinfo{note}{The exponent $\eta$ in Refs.~\cite{guo:96} and \cite{rieger:96b}
  is our $\eta + z$. Also the value of $\eta$ in Ref.~\cite{guo:96} is actually
  $1.1$, not $0.9$, as noted by Ref.~\cite{read:95}.}
\end{thebibliography}

\end{document}